\documentclass{article}
\usepackage{fa2020}
\usepackage{amsmath}
\usepackage{cite}
\usepackage{url}
\usepackage{graphicx}
\usepackage{color}
\usepackage[binary-units=true]{siunitx}

\title{Clarity: Machine Learning Challenges\\ to Revolutionise Hearing Device Processing}

\multauthor
{Simone Graetzer$^1$ \hspace{1cm} Michael Akeroyd$^2$ \hspace{1cm} Jon Barker$^3$ \hspace{1cm} Trevor J. Cox$^1$ } { \bfseries{John F. Culling$^4$ \hspace{1cm} Graham Naylor$^5$ \hspace{1cm} Eszter Porter$^2$ \hspace{1cm} Rhoddy Viveros Mu\~{ n}oz$^4$}\\
 $^1$ Acoustics Research Group, University of Salford, UK\\
 $^2$ Hearing Sciences, Division of Clinical Neuroscience, School of Medicine, University of Nottingham, UK\\
 $^3$ Department of Computer Science, University of Sheffield, UK\\
 $^4$ School of Psychology, Cardiff University, UK\\
 $^5$ School of Medicine, University of Nottingham, UK\\
{\tt\small s.n.graetzer@salford.ac.uk, j.p.barker@sheffield.ac.uk}
}

\begin{document}

\maketitle
\begin{abstract}
In the Clarity project, we will run a series of machine learning challenges to revolutionise speech processing for hearing devices. Over five years, there will be three paired challenges. Each pair will consist of a competition focussed on hearing-device processing (``enhancement") and another focussed on speech perception modelling (``prediction"). The enhancement challenges will deliver new and improved approaches for hearing device signal processing for speech. The parallel prediction challenges will develop and improve methods for predicting speech intelligibility and quality for hearing impaired listeners.

To facilitate the challenges, we will generate open-access datasets, models and infrastructure. These will include: (1) tools for generating realistic test/training materials for different listening scenarios; (2) baseline models of hearing impairment; (3) baseline models of hearing-device processing; (4) baseline models of speech perception and (5) databases
of speech perception in noise. The databases will include the results of listening tests that characterise how hearing-impaired listeners perceive speech in noise. We will also provide a comprehensive characterisation of each listener’s hearing ability. The provision of open-access datasets, models and infrastructure will allow other researchers to develop
algorithms for speech and hearing aid processing. In addition, it will lower barriers that prevent researchers from considering hearing impairment.

In round one, speech will occur in the context of a living room, \textit{i.e.}, a moderately reverberant room with minimal (non-speech) background noise. Entries can be submitted to either the enhancement or prediction challenges, or both. We expect to open the beta version of round one in October for a full opening in November 2020, a closing date in June 2021 and results in October 2021. 

This Engineering and Physical Sciences Research Council (EPSRC) funded project involves researchers from the Universities of Sheffield, Salford, Nottingham and Cardiff in conjunction with the Hearing Industry Research Consortium, Action on Hearing Loss, Amazon, and Honda. To register interest in the challenges, go to  \url{www.claritychallenge.org/}.
\end{abstract}

\section{Hearing loss and the problem of speech in noise}
Speech communication often takes place in noisy environments. The effects of noise on speech recognition vary according to the listener's hearing ability, the sources of noise, and the number and types of talkers. The most common complaint from listeners with hearing loss is difficulties communicating with others. 

One in six people in the UK has some level of hearing loss \cite{AoHL}, and this number is certain to increase as the population ages. Yet only 40\% of people who could benefit from hearing aids have them, and most people who have the devices don't use them regularly \cite{AoHL}. A major reason for this low uptake and use is the perception that hearing aids perform poorly. Despite the importance of speech-in-noise understanding to hearing aid wearers, speech in noise is still a critical problem, even for the most sophisticated devices. A hearing aid wearer may have difficulty when conversing with family or friends while the television is on, and when hearing public announcements at the train station. Such difficulties can lead to social isolation, and thereby reduce emotional and physical well-being.

Our approach to the problem of hearing aid processing of speech in noise is inspired by the latest developments in automatic speech recognition and speech synthesis, two areas in which public competitions have led to rapid advancements in technology. We want to encourage more researchers to consider how their skills and technology could benefit the millions of people with hearing impairments.

\begin{figure}[h]
 \centerline{\framebox{
 \includegraphics[width=\columnwidth]{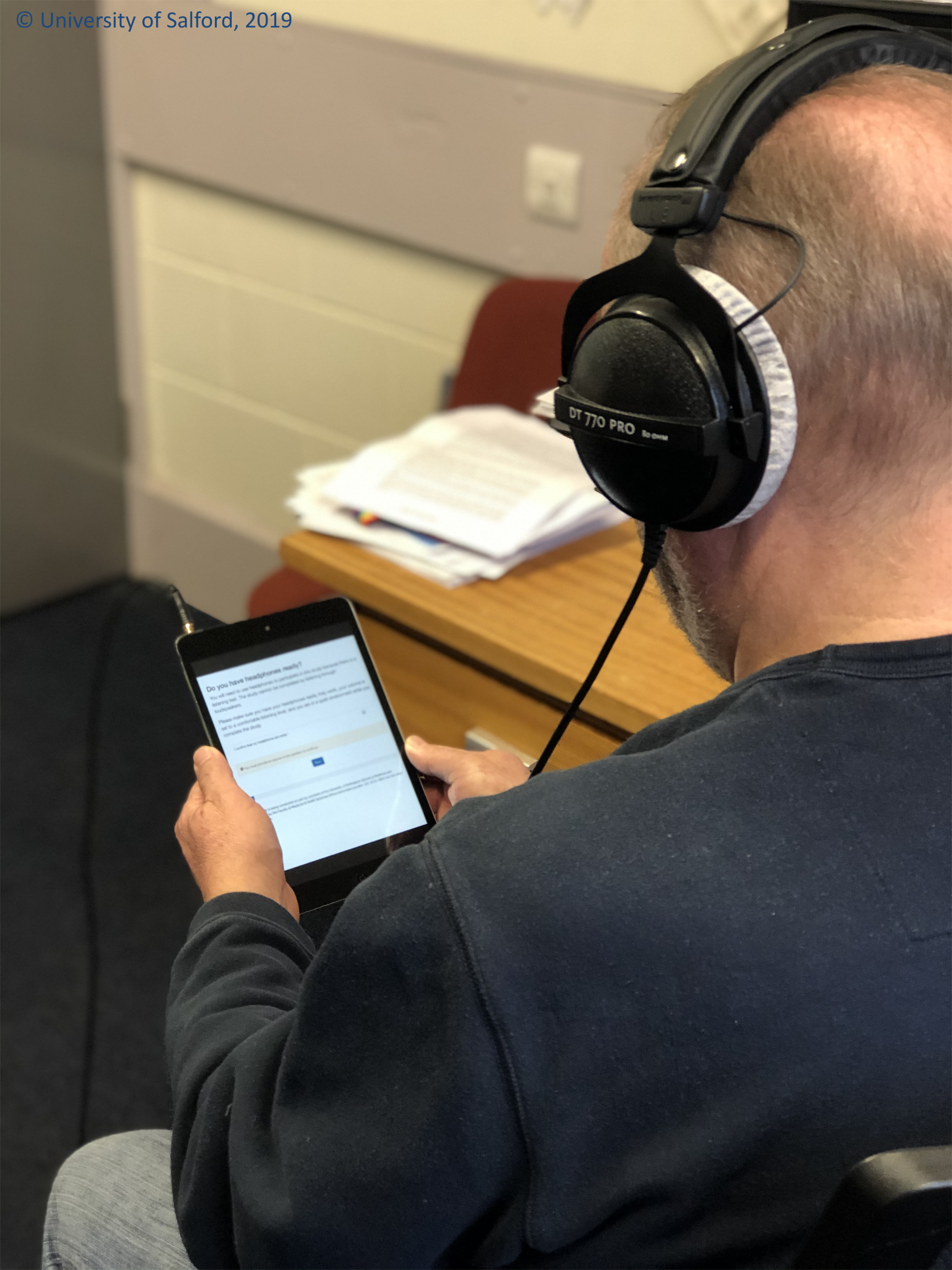}}}
 \caption{The listener will listen at home to speech signals transmitted via a tablet and headphones (\textcopyright University of Salford).}
 \label{fig:listen}
\end{figure}

\section{Advancing hearing aid signal processing via public competition}
In fields such as automatic speech recognition, researchers have developed algorithms that could improve speech intelligibility or clarity. Rapid advances are now being made in source separation, acoustic modelling and language modelling, that are driven by big data and deep learning. These novel techniques are too computationally-demanding for traditional hearing devices, but could be exploited by a new generation of powerful mobile devices that employ low-latency wireless technology. 

The Clarity project is inspired by those challenges that have accelerated many other fields of speech technology. Much of the recent advancement in the field of automatic speech recognition has been due to competitions such as CHiME, which is currently in its 6$^{th}$ iteration \cite{CHiME-6}. In 2012, PASCAL CHiME \cite{Barker} was designed to bring together researchers from various communities to foster novel approaches to the problem of ASR in reverberant environments with multiple dynamic noise sources. This is a task that corresponds to the speech-in-noise task faced by humans in everyday listening. Some components of the CHiME ASR solutions, such as source separation and de-reverberation, are directly relevant to hearing device processing but, until now, have been unevaluated in a hearing context.

We want to create resources that are as adaptive and generalisable to different domains as the CHiME baseline software, and to make the Clarity challenges as self-sustaining beyond the life of the original funding, so that we can generate long-term impact in a variety of real-world application areas. 

\section{Clarity data and scene specification}
\begin{figure*}[h]
 \centerline{\framebox{
 \includegraphics[width=\textwidth]{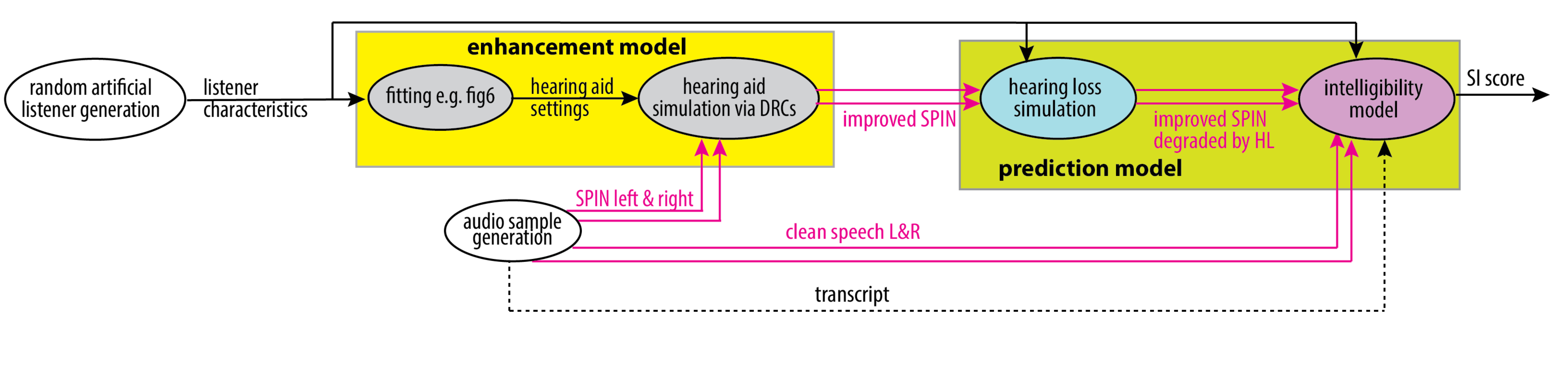}}}
 \caption{Baseline schematic for simulated listener data. In bright yellow is the clarity enhancement model and in mustard yellow is the clarity prediction model. Audio signals, such as Binaural SPeech In Noise (SPIN) signals, are shown in bright pink. The hearing loss (HL) model is shown in light blue. The output of the enhancement and prediction models is a set of Speech Intelligibility (SI) scores. Transcripts are not used by the baseline but will be provided to entrants.}
 \label{fig:pipeline}
\end{figure*}

In \figref{fig:pipeline} is shown an illustration of a baseline pipeline for simulated listener data and speech in noise signals created by means of the generative tool we will provide to competition entrants. A given set of listener characteristics and binaural spatialised speech in noise signals will be fed as input into the baseline hearing aid (clarity enhancement) model. The output will be fed into the clarity prediction model, which comprises a hearing loss model and a speech intelligibility model, which also receives as input the clean speech signals recorded in anechoic conditions and the matching transcripts of what was said. The output is the speech intelligibility scores for the hearing aid processed signals.

In the round one scenario, the speech signals, which are sentences from the British National Corpus \cite{BNC} produced by British English speakers in anechoic conditions, are emitted from a single source in a cuboid ``living room”. The receiver position is randomly selected over the area at least one metre from the source and the walls. The receiver, who is facing the source, is wearing hearing aids with either two or three microphones per ear. Reverberation time is low to moderate. There are multiple, predominantly non-speech, interferers, such as the television.

After undergoing a comprehensive hearing assessment in house, fifty listeners with healthy and impaired hearing will listen to the speech signals in their own home via a tablet and headphones (\figref{fig:listen}). They will listen to each sentence and respond by reproducing the sentence verbally. From these responses, speech intelligibility scores will be calculated. 

\section{Clarity competition rules}
Entries to the enhancement challenge should be causal, \textit{i.e.}, not using any sample information more than 5ms into the future. Entries will be be ranked according to mean intelligibility score across all audio signals in the test set. We will choose which systems will be evaluated by the listener panel based primarily on these rankings. Entries to the prediction challenge will be evaluated according to the mean-squared error between predicted speech intelligibility and measured speech intelligibility (from the listening panel) for all audio signals in the test set. For both challenges, only two entries from any one team can be evaluated by our listener panel. Teams must provide a two page technical document describing the system and any external data and pre-existing software used. Anonymous entries will not be eligible for cash prizes. Teams will be encouraged but not required to provide access to the system. There are no limits on computation costs, the amount of training data that can be generated, or the number of institutions involved in an entry. Teams may choose not to use the provided baseline models but must use the provided audio files and signal generation tool.

\section{Clarity outcomes}
The aim of the project is to improve the lives of many of the millions of people in the UK who have a hearing impairment. We will do this by making hearing aids more effective in processing speech in noise. 

The scientific legacy of the project will be improved algorithms for hearing aid processing, a test-bed that readily allows further development of algorithms, and more consideration by speech researchers of the hearing abilities of the whole population. The outcomes of the challenges will include new open-source software for large-scale generation of spatialised speech in noise signals, open-source hearing loss, hearing aid, and speech intelligibility algorithms, a new database of British English speech recordings in anechoic conditions, and tools to generate artificial listener characteristics. 

The full opening of round one will occur in November 2020. To be kept up to date on developments in the project, sign up at \url{http://claritychallenge.org/sign-up-to-the-challenges}.


\bibliography{FA2020_Clarity_Graetzer}

\end{document}